\documentclass[twocolumn,aps,prb,superscriptaddress]{revtex4-2}
	\usepackage{chngcntr}
	\usepackage{graphicx}
	\usepackage{color}
	\usepackage{amsmath}
	\usepackage{enumitem}
	\usepackage{amssymb}
	\usepackage{cancel}
	\usepackage{ulem}
	\usepackage{multirow}
    \usepackage[colorlinks=true,linkcolor=blue,citecolor=blue]{hyperref}

	\newcommand{\be}{\begin{equation}}
		\newcommand{\ee}{\end{equation}}
	
	\newcommand{\bea}{\begin{eqnarray}}
		\newcommand{\eea}{\end{eqnarray}}

	\newcommand{\lb}{\left[}
	\newcommand{\rb}{\right]}
	\newcommand{\lp}{\left(}
	\newcommand{\rp}{\right)}

	\renewcommand{\vec}[1]{{\boldsymbol #1}}

\usepackage[dvipsnames]{xcolor}

	\usepackage{bbold}

\begin{document}

\title{A controlled expansion for pairing in a polarized band with strong repulsion
%:  application to multi-layer graphene. %motivated by tetralayer graphene
}
%Spin-dependent Berry phase %, Landau levels 
%Geometric phases, flat bands and enhancement of superconductivity in chiral metals}%Spin-dependent geometric gauge fields and} enhancement of superconductivity in chiral metals \sout{due to spin chirality}}
\author{Zhiyu Dong}
\affiliation{Department of Physics and Institute for Quantum Information and Matter, California Institute of Technology, Pasadena, California 91125}
\author{Patrick A. Lee}
\affiliation{Department of Physics, Massachusetts Institute of Technology, Cambridge, MA 02139}

\begin{abstract}
Can strong repulsive interactions be shown to give rise to pairing in a controlled way? We find that for a single flavor polarized band, there is a small expansion parameter in the low density limit, once the Bloch wavefunction form factor is taken into account. A perturbative expansion is possible, even if the interaction is much stronger than the Fermi energy $\epsilon_F$. As a matter of principle, our work shows analytically how strong pairing can emerge from strong repulsion.  We illustrate our method with two examples: a 2D Dirac model and a 1D tight binding model with two orbitals. In the latter case, using density matrix renormalization group, we show that the analytical theory indeed  guided  us to discover the parameter regime where p-wave pairing with order-1 strength is dominant. 
%that is often used to describe multi-layer rhombohedral graphene and comment on the relationship with experiments.% This work opens a reliable pathway to search for topological superconductors with high-$T_c$ (relative to $\epsilon_F$) in materials with strong interactions. We summarize the requirements and suggest some illustrative design structures.
\end{abstract}
\maketitle

%\begin{widetext}
%\addZ{Significance statement: Whether purely repulsive electron-electron interactions can generate high temperature superconductivity is a longstanding question. Many candidates have been discovered in a variety of experimental systems, but mostly they are in the strongly repulsive regime where conventional perturbative approaches are uncontrolled and analytic theories are not available. Here we establish a controlled expansion valid for polarized bands with strong repulsion, revealing a reliable pathway for pairing to occur. Guided by this analytic framework, we predict $p$-wave pairing in a simple two band model, as confirmed by accurate numerical methods in one dimension. This success shows that our theory can provide a systematic foundation for discovering new high-Tc and topological superconductors. }
%\end{widetext}

The observation of high-$T_c$ superconductivity in various systems characterized by strong electron–electron repulsion raises a fundamental theoretical question: can purely repulsive electron–electron interactions alone give rise to pairing? The answer is positive according to numerical studies on models such as the 2D t-J model (for a recent paper which contains earlier references, see \cite{chen2025global}) and recently Landau levels in the presence of a periodic potential \cite{wang2025chiral}, but this occurs in the strong coupling limit where analytic treatments are not controlled. A perturbative approach was provided by the Kohn–Luttinger (KL) mechanism: this early work  \cite{Kohn1965} made use of the $2k_F$ singularity to show that in 3D there is attraction in some high angular momentum channel $l$ in second-order perturbation theory which can overwhelm the first-order repulsion for some large $l$. %This scenario was later extended to the $textit{l}=1$ (p-wave) case by Fay and Layzer \cite{Fay1968}, predicting a Tc consistent with measurements in $^3$He \cite{Osheroff1972}. Subsequently, Chubukov extended the framework to two-dimensional systems by incorporating higher-order effects \cite{Chubukov1993}. 
There have been notable extensions \cite{Chubukov1993,baranov1993enhancement,kagan1991strong,Maiti2013,Kagan2016,Kagan2014} 
%Saurabh Maiti ? Kagan review?
, and reformulation using the modern language of renormalization group (RG)  \cite{Zanchi1996,Zanchi2000,Raghu2010}
. However, the perturbative nature of this method means that the transition temperature is very low, with $T_c$ that is exponentially small in  $-l^4$ in the original KL theory which has been carried out to second order in the coupling constant. We note that extensions to include higher order terms can lead to a somewhat different conclusion.  In the case of low density, Fay and Layzer\cite{Fay1968} showed that the leading pairing channel is $l=1$. A recent re-examination of the RG approach \cite{fujimoto2025renormalization} have found that inclusion of higher order diagrams gives a $T_c$ that scales exponentially with $-l$ rather than $-l^4$. We also mention that  specially designed repulsive Hamiltonians have  been proven to exhibit pairing.\cite{Slagle2020,Crepel2021}
%However, strictly speaking, KL theory is asymptotically valid only in the weak-coupling regime, $g \ll 1$. In contrast, many realistic high-$T_c$ superconducting platforms exhibit rich charge and spin orders, with superconductivity either emerging on top of these orders or coexisting alongside them. The presence of such orders strongly suggests the SC occurs in a regime of strong coupling $g\gtrsim1$.
Nevertheless, a controllable theoretical framework that remains valid in the strong-coupling regime and results in high $T_c$ is highly desirable. This is the goal of the current paper.

In attempting to reach strong coupling, a common approach is to use the random phase approximation (RPA)\cite{Scalapino2012}, which sums over certain geometrical series and ignores other diagrams. In its simplest version (e.g. when applied to spinless or spin-polarized systems), the RPA essentially keeps track of the screening of charge-charge repulsion between electrons. The screened repulsion $V= \frac{V_0}{1+V_0\Pi}$ where $\Pi(q,\omega)$ is the polarization bubble, saturates at $1/\Pi$  for strong repulsion. This screened interaction therefore gives an order-1 coupling constant, and its frequency and momentum dependence has been used perturbatively to yield high $T_c$. The RPA approach has been applied to cuperates\cite{Scalapino1986, Bickers1989,Millis1990,Monthoux1991,Bulut1992,Scalapino1995} and recently to tetralayer and pentalayer graphene\cite{chou2024intravalleyspinpolarizedsuperconductivityrhombohedral,geier2024chiraltopologicalsuperconductivityisospin,yang2024topologicalincommensuratefuldeferrelllarkinovchinnikovsuperconductor,qin2024chiralfinitemomentumsuperconductivitytetralayer,jahin2024enhancedkohnluttingertopologicalsuperconductivity,Parra-Martinez2025} and partially filled Chern bands. \cite{Guerci2025}%where several SC phases are found on top of a flavor-polarized background\cite{han2025signatureschiralsuperconductivityrhombohedral}.

%However, the justification of RPA is questionable in many settings. The procedure of focusing on screening diagrams is somewhat justification when there are $\mathcal{N}\gg1$ flavors of electrons due to a small parameter $1/\mathcal{N}$. However, it is no longer justified for systems with small values of $\mathcal{N}$: there is no good reason to single out the screening processes while neglecting other processes such as vertex corrections, cross diagrams, etc. An extreme example is the case of flavor-polarized ($\mathcal{N}=1$) bands where, at each order of the coupling constant expansion, these types of diagrams should have comparable contributions. Therefore what RPA accounts for is only a small part of the whole story. 

However, the justification of RPA is questionable. When there are $\mathcal{N}\gg 1$ electron flavors,  a small parameter $1/\mathcal{N}$ provides some control. Yet, most systems does not have large $\mathcal{N}$, and this justification breaks down: there is no  reason to ignore vertex corrections or crossed diagrams. An extreme case is that of flavor-polarized ($\mathcal{N}=1$) bands. Take the example of a short range delta function repulsion. Pauli exclusion tells us that this interaction does nothing, while RPA gives a coupling $1/\Pi(q,\omega)$ in the strong coupling limit. Clearly any pairing arising from the $q$ and $\omega$ dependence of this coupling is an artifact in the delta function limit. Consequently, there are reasons for caution when the interaction has short but finite range.  %In this paper we turn this situation around by taking advantage of the simplicity of the short range repulsion limit of the flavor polarized  problem as use it as the starting point of an expansion that is controllable
%strongly-interacting flavor-polarized electrons might not be that difficult. 
%In this paper, we show that there is a new small parameter that can make the perturbation approach controllable even for spontaneously flavor-polarized electrons in which the original coupling is strong.
%even when the bare e-e interaction is strong. Using this theory, we demonstrate that superconductivity with an order-1 pairing interaction strength can be reliably achieved. %and yields predictions for p-wave superconductivity emerging from pure repulsion in the spin-polarized band. 
%In this paper, 

In this paper we turn this situation on its head by taking advantage of the special property of a fully polarized band: Fermi statistics keep electrons apart, so that a short range delta function repulsion has no effect. By slightly relaxing the delta function we show that there is a small expansion parameter in the low density limit, when the Fermi momentum $k_F$ is much less than a characteristic momentum scale of the repulsive interaction%the inverse lattice spacing
. By including the Bloch wavefunction, we show that a substantial superconducting $T_c$ can be calculated in a controlled way. 
%We will then apply this approach to multilayer graphene to predict a SC phase diagram as in Fig.\ref{fig:pocket size}, which resembles the measurements in Ref.\cite{han2025signatureschiralsuperconductivityrhombohedral}.

After presenting the  theoretical formulation, we describe two examples to illustrate our method. The first is an N-Dirac model that has been widely used as an approximate description of N layer rhombohedral graphene, and the second is a simple tight binding model with two orbitals per unit cell. In the latter case we predict strong pairing in a certain parameter range and we confirm our prediction in a 1D version using DMRG.

\begin{figure}
    \centering
    \includegraphics[width=0.98\columnwidth]{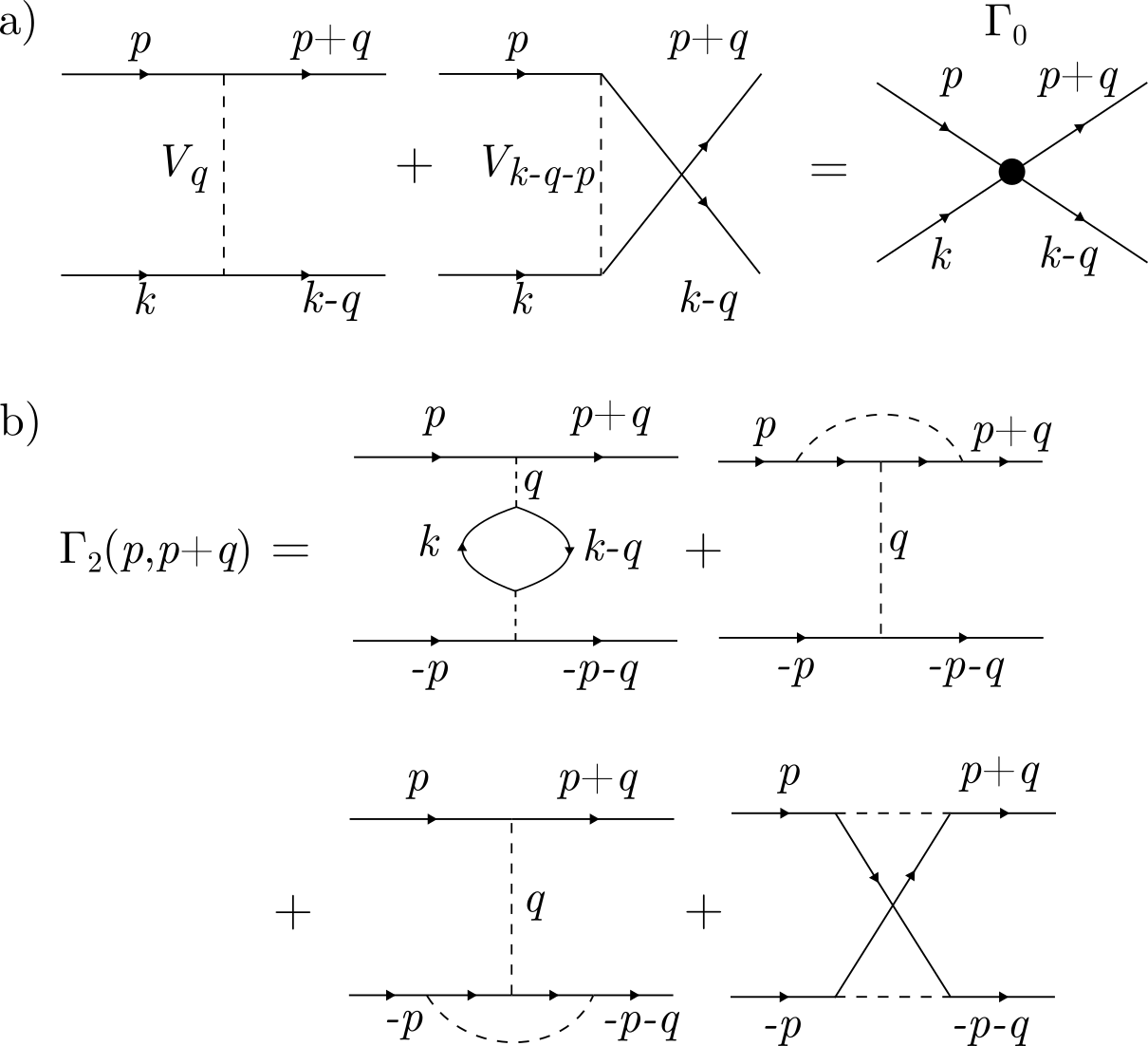}
    \caption{a) The first order diagram consists of direct repulsion and exchange. The total scattering amplitude $\Gamma_0$ (the filled circle) is much smaller than each single diagram due to the weak momentum dependence of the wavefunction and $V_q$ (see text around Eq.\eqref{eq:Gamma0} and Eq.\eqref{eq:Gamma0=alpha^2}). The 1st-order pairing interaction $\Gamma_1$ can be obtained by setting $k$ to be $-p$ in the first diagram in a). Here, $k =(\omega,\vec k)$. b) The second-order pairing interaction $\Gamma_2$ is the sum of four diagrams (bubble, vertex correction and cross diagrams).}
    \label{fig:diagrams}
\end{figure}

To start, we consider a charge-charge repulsion in fully  flavor-polarized electrons:
\be
H_{int} = \sum_{\vec q} \frac{V_{\vec q}}{2} :\rho_{\vec q} \rho_{-\vec q}:, \quad \rho_{\vec q} = \sum_{\vec k} \Lambda_{\vec k-\vec q, \vec k} c^\dagger_{\vec k-\vec q} c_{\vec k}.
\ee
where the form factor $\Lambda_{\vec k', \vec k}= \langle u_{\vec k'}|u_{\vec k}\rangle $, $|u_{\vec k}\rangle$ represents the Bloch wavefunction. Here we have neglected  umklapp processes; the rationale will be discussed later.
We start from a simplest case of electrons in tight-binding one-band model with a contact interaction, so that the Bloch wavefunction and interaction are both independent of momentum:
i.e. $|u_{\vec k}\rangle = 1$ and $V_{\vec q} =V_0$. For this case, one finds the interaction effect completely vanishes through a perfect cancellation between direct and exchange processes (see Fig.\ref{fig:diagrams}(a) that always come in a package at any order of diagrammatic expansion. This enforces the Pauli exclusion which does not allow electrons in one flavor to interact through a contact interaction.

This fact motivates us to consider %whether some seemingly unsolvable strongly-coupled system can be viewed as a ``small" twist on top of
a  modification of this trivially solvable case in two steps.%, thus allowing a controllable perturbative theory for strongly-coupled flavor-polarized electrons.
 
\begin{enumerate}
\item[(1)]{We introduce some momentum-dependence by truncating the repulsion $V_{\vec q}$ at a momentum $q_d$ such that $k_F\ll q_d \ll G_0$, where $G_0$ is the shortest reciprocal lattice vector i.e. 
\be\label{eq:V(q)}
V_{\vec q} = V_0\Theta(q_d-|\vec q|).
\ee
}
This can be realized by an adjacent metallic gate at a distance $d=1/q_d$.
\item[(2)]{With Eq. \ref{eq:V(q)}, the interaction still exactly cancels if the Bloch function $u_{\vec k}$ is independent of ${\vec k}$, as in the case of a tight-binding band with a single orbital.
 A non-vanishing effect comes from the $k$-dependence in the Bloch wave function for the momenta of interest, %, ``weak" in the sense that the amplitude of $k$-dependent part of wavefunction is much smaller than the $k$-independent part. 
 which is present for a general LDA type band or a  multi-band tight binding Hamiltonian. The strength of this matrix element effect depends on details such as proximity to a hybridization gap or  Berry curvature.}
\end{enumerate}
%To gain some insights, 
%This certainly makes the model more realistic, 
%such as in multilayer graphene which we will comment on later. 
%but can it be controllably treated through perturbation approach? The answer is positive. To see this, we inspect the two steps closely:

%Let's track the change after each step of manipulation:
 %It is important to note that Step(1) is not really a ``small" modification of the interaction in that it turns a delta function repulsion on the sub-lattice scale to one that extends over several unit cells and all high order umklapp processes needed to recover the delta function interaction have been ignored. 
 We note that in a lattice model, in order to capture the delta function repulsion on the sub-lattice scale, it is necessary to include all umklapp terms. Therefore, it is not surprising that the exclusion of umklapp terms in our model can lead to nontrivial effect. On the other hand, in a model where the range of repulsion is smaller than $1/k_F$, ignoring umklapp has almost no impact on the low-energy physics. This is because the scattering processes that are relevant for low-energy physics are those with all the electron's momenta restricted within $O(k_F)$. 
 %which we assume throughout our analysis to be $\ll q_*$ %\footnote{Here we assume the band dispersion is steep enough (or open a large gap) at $k> q_*$ which is the situation we will focus on}. As a result, exchange and direct diagram in Fig.~\ref{fig:diagrams}a) still exactly cancel each other, so electrons are still effectively noninteracting. 
%One can alternatively understand it from r-space consideration: For contact interaction, electrons are hardball of radius $a$ (atomic scale), now they become hard balls of $1/q_*$ which is still much smaller than their average distance $1/k_F$. So it is not surprising that this manipulation does not affect low-energy physics. Therefore, after \textit{Step1}, the interaction still does nothing to low-energy physics.
%As in the original model interaction does not do anything, after first step the interaction still does nothing.

With Step(2), the cancellation between exchange and direct diagrams in Fig.~\ref{fig:diagrams}a) becomes imperfect. Specifically, the total scattering amplitude of two electrons at $\vec k,\vec p$ to $\vec k+\vec q$ and $\vec p-\vec q$ ($|\vec k|,|\vec p|,|\vec q| \lesssim k_0$) through direct and exchange processes is given by
\be\label{eq:Gamma0}
\Gamma_0(\vec k,\vec p,\vec q) =
V_0\lb \Lambda_{\vec k,\vec k+\vec q}%\langle u_k|u_{k+q}\rangle 
\Lambda_{\vec p,\vec p-\vec q}%\langle u_{k'}|u_{k'-q}\rangle
- 
\Lambda_{\vec k,\vec p-\vec q}%\langle u_k|u_{k+q}\rangle 
\Lambda_{\vec p,\vec k+\vec q}%\langle u_{k'}|u_{k'-q}\rangle
%= V_0 \lb \langle u_p|u_{p+q}\rangle\langle u_{k}|u_{k-q}\rangle -  \langle u_p|u_{k-q}\rangle\langle u_{k}|u_{p+q}\rangle \rb
\rb
\ee
This amplitude $\Gamma_0$ is nonvanishing and comparable to $V_0$ for generic $\vec k$-dependent $|u_{\vec k}\rangle$. However, for dilute electrons, this non-vanishing total scattering amplitude $\Gamma_0$ is controlled by a new small parameter.  To see this explicitly, we use a two band model where the Fermi level lies in one band which is %flat and 
separated by an energy gap from the other band. %to the case where the flat region of a band which has a small radius of $k_0$ that is comparable to the fermi momentum $k_F$. We study processes in Fig.\ref{fig:diagrams} but with all momenta restricted within $O(k_F)$, as these are the processes relevant for low-energy physics. 
To estimate the amplitude $\Gamma_0$, we express the Bloch wavefunction in terms of its constant part and $k$-dependent part: $|u_{\vec k}\rangle  = \sqrt{1-|\alpha_{\vec k}|^2}|v_0 \rangle + \alpha_{\vec k} |v_1\rangle$, where $|v_0\rangle$ represents an ``typical" Bloch wavefunction; $|v_1(\vec k)\rangle$ is orthogonal to $|v_0\rangle$,% representing the variation of the Bloch function; 
$|\alpha_k|$ represents the magnitude of  the Bloch function's variation, which is small for $k\lesssim O(k_F)$ due to the smallness of $k_F$. Assuming the wavefunction has an order-1 variation on some momentum scale $k_*$, 
%on the scale of the shortest reciprocal lattice vector $G_0$, 
the parameter $|\alpha_{k_F}|$ is small in $k_F/k_*$ ($k_*$ is expected to be $G_0$ or larger for a generic band, and  is roughly $G_0$ for a generic Chern band. (It is of order the inverse of the magnetic length $l_B$ for a Landau level.) Alternatively, the new  parameter $|\alpha_k|$ can be expressed in terms of 
the quantum metric $g_{ij}(k)$ as roughly $\sim \frac1{A} \int_A d^2\vec k g_{ij}(0) k_ik_j $, where $A$ is the relevant k-space area. In the presence of a Berry curvature, $g_{ij}$ is not small and we rely on the smallness of $k_F$ to get a small parameter. More generally, $g_{ij}$ depends on details of the  Bloch band. 
Plugging the expression of wavefunction  $|u_k\rangle$ 
%form of wavefunction 
into Eq.\eqref{eq:Gamma0} we find
\be\label{eq:Gamma0=alpha^2}
\Gamma_0(\vec k,\vec p,\vec q)
%V_{\vec q} 
%\Lambda_{\vec k,\vec k+\vec q}%\langle u_k|u_{k+q}\rangle 
%\Lambda_{\vec k',\vec k'-\vec q}%\langle u_{k'}|u_{k'-q}\rangle
%- V_{\vec k'+\vec q-\vec k}
%\Lambda_{\vec k,\vec k'-\vec q}%\langle u_k|u_{k+q}\rangle 
%\Lambda_{\vec k',\vec k+\vec q}%\langle u_{k'}|u_{k'-q}\rangle
\sim O(|\alpha_{k_F}|^2 V_0 )
\ee
which can be much smaller than $V_0$.
%which can be much weaker than $V_0$ due to the small parameter $\alpha$.
Therefore, the effective strength of coupling in this system is described by the typical value of the total vertex $\Gamma_0$ (rather than the original vertex $V_0$) on the Fermi surface: 
\be\label{eq:g_eff}
g_{\rm eff}= \nu_0 \int \frac{d \theta_{\vec p}d \theta_{\vec p'}}{(2\pi)^2} \Gamma_0(\vec p,\vec p’,0),
\ee
where $\theta_{\vec p}$ is the polar angle of $\vec p$, the integrals over $\vec p, \vec p'$ are taken along Fermi surface and $\nu_0$ is the density of states. Even if the original coupling $g_0 = \nu_0V_0\gg 1$, there still exists a regime of $\alpha_{k_F}$ such that the effective coupling constant $g_{\rm eff}$ is still sufficiently weak so that a 
controlled analysis through a perturbative expansion is possible.
%in terms of $\alpha_{k_F}$.

%Below we consider pairing by solving the gap equation perturbatively in $\Gamma_0$. A key point is that  the first order contribution, which is typically repulsive, is exactly cancelled for the interaction given in Eq. \ref{eq:V(q)} provided   the Bloch function has inversion symmetery $|u_{\vec k}\rangle = |u_{-\vec k}\rangle $. The leading contribution is second order in $\Gamma_0$. Since second order perturbation  is negative, we can expect to find an attractive channel for pairing. In this way we avoid the need to find a way to overcome the leading repulsive term as in the case for the Kohn-Luttinger mechanism which is responsible for the exponentially small  $T_c$ , as mentioned earlier.

The reader may be concerned that the bare interaction remains strong and may appear in other diagrams outside of the pairing channel considered below. Here we appeal to Landau's Fermi liquid theory which states that  interaction effects, no matter how strong, that are far away from the Fermi surface only give rise to renormalized parameters such as effective mass and coupling strength for the low energy quasi-particles. Hence our treatment only deals with pairing of quasi-particles in the Landau sense and the bare interactions are treated as renormalized parameters which remains strong. While the general framework remains valid, this renormalizaion may play an important role if one attempts to predict $T_c$ starting with a microscopic Hamiltonian. We have to keep this in mind in our choice of microscopic models, as  further discussed below.
%Scattering processes relevant for low-energy physics are those with all initial and final momenta within the flat region of the band. 
%In superconducting regime in realistic graphene multilayers,
%For simplicity, we set the flat-region radius $k_0$ in this mdoel to be comparable to $k_F$. 
%So below we will use Fermi momentum $k_F$ to refer to this momentum scale. 
%The Bloch wavefunction of states inside this momentum scale are usually highly sublattice-polarized, so that $\alpha_{k_F} \sim O(h_1(k_F)/h_3)\ll 1$.  Using these, we find the majority contribution to $\Gamma_0$, which is the momentum-independent part, is canceled out, so that the total scattering amplitude $\Gamma_0$ is much weaker than the strength of direct repulsion $V_0$:
%\be\label{eq:Gamma0/V0}
%\Gamma_0(k,k',q) /V_0 \sim  O(\alpha_{k_F}^2) \ll 1
%\ee

%All diagrams in series expansion gradually shows up in this step, but so long as $\alpha_{k_F}$ is small, series is well controlled
The method described so far is generally applicable for any band structure. Below we demonstrate this idea through two concrete examples: an $N$-Dirac model and a two orbital tight binding model. 

\textit{Dirac model:}
First we consider an $N$-Dirac model with following noninteracting continuum Hamiltonian
\be
H_0(\vec p) = 
\lp 
\begin{matrix}
    \frac{u}{2} &  \frac{u\lp p_x-ip_y\rp^{N}}{2k_0^N} \\
    \frac{u\lp p_x+ip_y\rp^{N}}{2k_0^N}  & -\frac{u}{2}
\end{matrix}
\rp
\ee
Here $N$ can take integer values $N=1,2,3,4...$ while $u$ sets the gap between two bands and flattened them. The momentum $k_0$ sets the radius of the flattened band bottom(top). For $|\vec p|$ exceeding this scale, the band dispersion becomes steep. Without losing generality, we  focus on the electron-doped case ($n>0$) throughout our analysis. The Bloch wavefunction in the electron band is given by $|u_{\vec k}\rangle = (\sqrt{1-|\alpha_{\vec k}|^2},|\alpha_{\vec k}| e^{iN\theta_{\vec k}})$ with $\theta_{\vec k}$ representing the angle of $\vec k$, $|\alpha_{\vec k}| \sim \frac1{2}(|\vec k|/k_0)^N$ for $|\vec k|\ll k_0$.
%Next, on top of this noninteracting Hamiltonian, we add the following short-range repulsive interaction:
%\be
%H_{int} = \frac{1}{2}\sum_{\vec q} V_{\vec q} \rho_{\vec q} \rho_{-\vec q}, \quad \rho_{\vec q} = \sum_{\vec k} \Lambda_{\vec k, \vec k+ \vec q} c^\dagger_{\vec k} c_{\vec k+\vec q}.
%\ee
%with $V_q$ taking the form of Eq.\eqref{eq:V(q)}.
We note in passing that this model is a widely used as a toy model for real systems such as rhombohedral graphene with N layer (see e.g.\cite{geier2024chiraltopologicalsuperconductivityisospin}). Remarkably, SC phases are indeed seen in flavor-polarized phases in some of these systems. \cite{han2025signatureschiralsuperconductivityrhombohedral} We should point out that this model does not fully describe the experimental system in that realistic features such as warping are ignored. In this paper we use this model as an illustration of our framework and
make no further discussion of its relation to the real system. 
Within our framework, the pair-scattering processes  in  first and second order in $V_0$ can be expressed as diagrams in Fig.\ref{fig:diagrams}. The first-order pair interaction is given by
\be\label{eq:Gamma1}
\Gamma_1(\vec p,\vec p') = V_0 \langle u_{\vec p}|u_{\vec p'}\rangle\langle u_{-\vec p}|u_{-\vec p'}\rangle 
%-  \langle u_p|u_{-p'}\rangle\langle u_{-p}|u_{p'}\rangle \rp
\ee
%Due to the rotational symmetry of our toy model, the Bloch wavefunction satisfies
While $\Gamma_1$ is comparable to $V_0$, the effective pairing interaction is implicitly small in $\alpha_{k_F}$. This is because, due to Fermion statistics, in $\Gamma_1$ only its antisymmetric part is useful for pairing. The antisymmetric part is given by, $\frac{1}{2} \lp \Gamma_1(\omega-\omega';\vec p,\vec p') - \Gamma_1(\omega+\omega';\vec p,-\vec p') \rp$, which is equivalent to the two diagrams in Fig.\ref{fig:diagrams}a), therefore has the same cancellation as $\Gamma_0$. 
The second-order pairing interaction is implicitly $\propto \Gamma_0^2$, and is thus expected to be higher-order in terms of the small parameter $\alpha_{k_F}$. %This is expected but may not be easy to see from Fig.\ref{fig:diagrams}b).
This is seen explicitly by noting that the sum of four diagrams in Fig.\ref{fig:diagrams}b) is equivalent to one bubble diagram with its two vertices replaced by two $\Gamma_0$'s.

Interestingly, for even-$N$ Dirac model ($N=2,4,6...$), the wavefunction has the following symmetry: $|u_{\vec p}\rangle = |u_{-\vec p}\rangle $. Therefore, the antisymmetrized pairing interaction %$\Gamma_1(\vec p,\vec p') = 
$\frac1{2}\lp \Gamma_1(\vec p,\vec p') -\Gamma_1(\vec p,-\vec p') \rp $ always vanishes. As a result, the pairing interaction in even-$N$ Dirac model is given by the second-order process (shown in Fig.\ref{fig:diagrams} b)) which is generally attractive. In contrast, for the odd-$N$ Dirac model ($N=1,3,5...$) the two first-order diagrams do not cancel each other, so $\Gamma_1$ is the leading order contribution. As a result, pairing is not expected to  occur as 1st order interaction usually disfavors pairing. %\addZ{We note in the passing that, for odd-frequency pairing channels which we will discuss briefly later, the 1st order interaction  cancel for both odd and even  %$\Gamma_1(\omega,p,\omega',p') = \Gamma_1(\omega,p,-\omega',p')$, 
 %$N$ due to the frequency independence of the bare interaction. }%\addZ{Therefore, from now on we solely focus on even-N Dirac models.}
%same cancellation works because symmetry enforces $|u_{p}\rangle = \tau_z |u_{-p}\rangle $, where $\tau_z$ is the Pauli matrix. 
%We note parenthetically that this cancellation of 1st-order processes generally holds true for all even-layer graphene (when trigonal warping is neglected), but not for odd-layer graphene.

%We comment on the distinction between our mechanism of pairing and Kohn-Luttinger(KL) scenario, as the fact that pairing interaction is dominated by 2nd-order diagrams in  even-$N$ Dirac model might sound similar to KL scenario\cite{Kohn1965}. In Kohn-Luttinger theory, bare e-e interaction is required to be weak. There, it is the behavior of polarization function below $2k_F$ that makes the second-order processes win over first-order ones in the large-angular-momentum ($l\gg1$) channel\cite{Kohn1965}.  In comparison, our theory is controllable even in a ``strong"-coupling regime (``strong" in the sense that the naive coupling constant $V_0\nu_0\gg1$), as we have the new small parameter $\alpha_{k_F}$ that controls the series expansion. In our setting, the second-order processes win over first-order ones because the first-order contribution to pairing interaction vanishes due to the form of wavefunction. The justification of our scenario does not rely on a large $l$.

%\textit{Pairing channels and $T_c$:}
Next, we proceed to analyze the pairing problem, focusing mainly on even $N$. 
%We will first study the pairing channel and the behavior of $T_c$ in Dirac model with a general integer value of $N$, and then obtain the $T_c$ in even-$N$ and odd-$N$ Dirac models separately. 
To start, we write down the linearized pairing gap equation:
\be\label{eq:gap eqn 1}
\Delta(\omega,\vec p) = T_c\sum_{\omega',\vec p'}\frac{\Gamma(\omega,\omega';\vec p,\vec p')\Delta(\omega',\vec p')}{\omega'^2 +\epsilon_{\vec p'}^2} 
%= \frac{1}{2} \sum_{\omega',p'}\frac{\lp \Gamma(\omega-\omega';p,p') - \Gamma(\omega+\omega';p,-p') \rp \Delta(\omega',p')}{\omega'^2 +\epsilon_{p'}^2}
\ee
 where $\Gamma$ is the two particle irreducible pairing interaction, whose first-order contribution is given in Eq.\eqref{eq:Gamma1} and second-order contribution is $\Gamma_2$ expressed diagrammatically in Fig.\ref{fig:diagrams}b). 
 %$\Gamma_{a}$ is the antisymmetrized pairing interaction : $\Gamma_{a}(\omega,\omega';\vec p,\vec p')= 
%\frac{1}{2} \lp 
%\Gamma(\omega-\omega';\vec p,\vec p') - \Gamma(\omega+\omega';\vec p,-\vec p') %\rp
%$, 
%At first order, $\Gamma(\omega,\omega';p,p') = \Gamma_0(p,-p,p'-p)$. 
To proceed, we neglect the radial momentum-dependence of $\Delta$ and integrating along the direction perpendicular to Fermi surface $p_\perp$. Reparameterizing momenta $\vec p,\vec p'$ using the angle $\theta, \theta'$ on Fermi surface yields:
\be\label{eq:gap equation theta}
\Delta(\omega;\theta) = \pi \nu_0 T_c\sum_{\omega'}\int \frac{d\theta'}{2\pi}\frac{\Gamma(\omega,\omega';\theta,\theta')\Delta(\omega';\theta')}{|\omega'|} 
\ee 
In our setting, $\Gamma(\theta,\theta')$ is a function of $\theta-\theta'$ as dictated by the U(1) symmetry of Dirac models (spatial rotation + a relative phase shift between AB sublattice). This allows labeling pairing channels with angular momenta $l$ which is a good quantum number:
\be\label{eq:gap equation angular momentum}
\Delta^{(l)}(\omega) = \pi\nu_0 T_c\sum_{\omega'} \frac{\Gamma^{(l)}(\omega,\omega')\Delta^{(l)}(\omega')}{|\omega'|}, \quad l\in \mathcal{Z}.
\ee 
Here we have defined 
%the projection of antisymmetrized interaction in channels: $\Gamma^{(l)}_a = \Gamma^{(l)} (\omega-\omega') - (-)^l \Gamma^{(l)} (\omega+\omega')$, and defined 
the partial wave components: $\Delta^{(l)}(\omega) = \int \frac{d\theta}{2\pi} \Delta(\omega;\theta) e^{il\theta}$ and
 $\Gamma^{(l)} (\omega-\omega') = \int \frac{d(\theta-\theta')}{2\pi} \Gamma (\omega-\omega';\theta-\theta') e^{il(\theta-\theta')}$. In Eq.\eqref{eq:gap equation angular momentum}, angular momentum $l$ is allowed to take either even or odd values. However, due to fermion statistics, the odd-parity pairing channels (i.e. channels with odd-valued $l$) have to be even in frequency, whereas the even-parity pairing channels have to be odd in frequency: $\Delta^{(l)}(\omega)= -\Delta^{(l)}(-\omega), l\in {\rm even}$.\cite{berezinskii1974new,RevModPhys.91.045005}
%As the gap equation should be worked out in quite different ways for even-frequency and odd-frequency channels, 
%Below we work out the $T_c$ of even-frequency and odd-frequency channels separately:
%$\Delta(\omega)$ needs to be even in $\omega$, when 

The $T_c$ of even-frequency odd-$l$ pairing is given by
\be
T_c^{(l)} \sim 1.13\epsilon_F \exp\lp -\frac1{g^{(l)}} \rp,  \quad l\in {\rm odd}
\ee
 which is similar to Tc in BCS problem, except that the bandwidth of pairing interaction, which is Debye frequency in BCS problem, is replaced with $\epsilon_F$. This is obtained through solving Eq.\eqref{eq:gap equation angular momentum} by replacing $\Gamma^{(l)}(\nu)$, which is an even function of frequency $\nu$ for $l \in {\rm odd}$, with $\Gamma^{(l)}(\nu) = \Gamma^{(l)}(0) \Theta(\epsilon_F-|\nu|)$ \footnote{This form is justified because the pairing interaction in our problem arises from 2nd-order diagram in Fig.\ref{fig:diagrams} which inherit the frequency dependence of the polarization bubble.} Here, we have defined the dimensionless coupling constant $g^{(l)} = \nu_0 \Gamma^{(l)}(0) $. As a reminder, we here only focus on even-$N$ Dirac models as only in these cases the first-order pair-breaking effect is canceled. Therefore, the pairing interaction purely arise from second-order contribution, so $\Gamma^{(l)} = \Gamma_2^{(l)}$, $g^{(l)} = g_2^{(l)}$ throughout our analysis below.
 
%\addZ{
%The Tc of odd-frequency pairing channel behaves differently. We analyze odd-frequency channel carefully in Appendix \ref{app:odd_frequency} and \ref{app:details of odd-frequency}. Here, without diving into details, we give a few qualitative expectation: First, as the odd-frequency pairing does not have the BCS logarithm\cite{RevModPhys.91.045005,berezinskii1974new}, we expect there to be a threshold coupling strength below which Tc vanished. Above this threshold, Tc becomes comparable to the Fermi energy. Meanwhile, we expect the presence of odd-frequency pairing is insensitive to $N$, as opposed to even-frequency ones. This is because, unlike even-frequency channels, the 1st order pairing interaction vanishes regardless of value of $N$ (see discussion above Eq.\eqref{eq:gap eqn 1}) for odd-frequency channels.  
%}

\begin{figure}
    \centering
    \includegraphics[width=0.98\columnwidth]{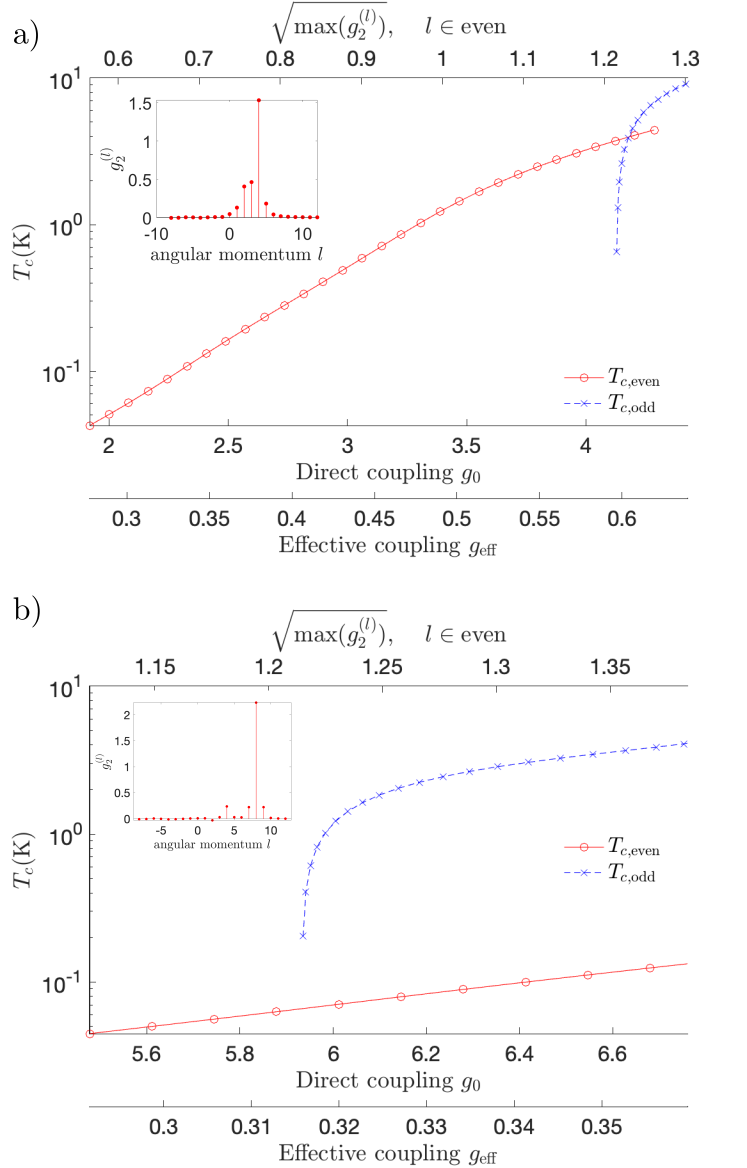}
    \caption{Dependence of $T_c$ on interaction strength for a) $N=2$ Dirac model and b) $N=4$ Dirac model. Red and blue curves represent $T_c$ in the leading even-frequency and odd-frequency channels respectively. In the bottom x-axes, we show the value of the bare and effective couplings $g_0$ and $g_{\rm eff}$ defined in Eq.\ref{eq:g_eff}
    .  %In the regime shown, although $g_0\gg 1$, the perturbation expansion is valid as $g_{\rm eff}\ll 1$. 
    The top x-axis shows the coupling strength in the leading odd-frequency pairing channel $\max(g^{(l)}_2)$ with $l$ restricted to be an even integer \cite{myfootnote1}.
    Results in 
    %a) is calculated at $n=4\times 10^{-3}{\rm nm}^{-2}$, $u=80{\rm meV}$, $k_0=0.4{\rm nm}^{-1}$, whereas results in b) is 
    both panels are calculated at $n=5\times 10^{-3}{\rm nm}^{-2}$, $u=80{\rm meV}$, $k_0=0.34{\rm nm}^{-1}$. The Fermi energy in two cases are $\epsilon_F = 5.51{\rm meV}$
    %1.94{\rm meV}$
    for a) and $\epsilon_F=1.75{\rm meV}$ for b). 
    In the regimes shown, despite the bare coupling $g_0\gg 1$, the effective coupling $g_{\rm eff}$ is below 1, so the perturbation theory is controlled.
    %, except that we should keep in mind that $T_c$ should be cutoff at $T_c\sim \epsilon_F$. 
    The insets show the angular momentum components of the pairing interaction. It is maximized at $l=4$ in $N=2$ Dirac model, and at $l=8$ in the $N=4$ one.}
    \label{fig:Tc}
\end{figure}

In Fig.\ref{fig:Tc} we present the numerical results of Dirac model, where panels a) b) describes N=2 and N=4 Dirac models separately. We find the leading even-frequency(odd-$l$) pairing channel is $l=3$ for $N=2$ Dirac model, and is $l=7$ for $N=4$ Dirac model. These are topological superconductors where the gap function goes as $e^{i l \theta}$. Their $T_c$ are shown as red curves in the figure. The insets in Fig. \ref{fig:Tc} (a,b) show the strength of 2nd-order pairing interaction in different partial wave channels. We find that $g^{(l)}$ are 
mostly positive, indicating attraction, and have a strong peak near $l=2N$. Its origin can be traced back to the fact that $\Gamma_0$ given by Eq. \ref{eq:Gamma0} has two factors of $\Lambda$. When all the momenta in the  $\Lambda$'s are on the Fermi surface, there is an $e^{iN\theta}$ contribution to $\Gamma_0$ where $\theta$ is the angle between $\vec k$ and $\vec k'$. To second order in $\Gamma_0$ we obtain a factor $e^{i2N\theta}$. This feature is specific to the $N$-Dirac model. The large peak for $g^{(l)}$ for even $l=2N$ does not contribute to conventional even frequency BCS pairing, which can make use of only of the largest odd $l$.  This motivates us to consider odd-frequency pairing  whose $T_c$ are shown as blue curves in Fig.\ref{fig:Tc}. There is no logarithmic singularity for odd-frequency pairing, hence a threshold in the effective interaction is required and $T_c$ becomes comparable to fermi energy. The requirement of strong coupling means that odd frequency pairing is not controllable and is included here only for completeness. (It is worth mentioning that for odd frequency  pairing the first order term cancels for both even and odd N  since this  relies only on the absence of frequency dependence of the bare interaction.) The analysis of odd-frequency pairing will be detailed in Appendix \ref{app:odd_frequency}. %\ref{app:details of odd-frequency} and \ref{app:check solution}. }

%This is reasonable because the pairing interaction is maximized for odd values of $l$  that are closest to $2N$. (see insets of Fig.\ref{fig:Tc}). 
%Nevertheless, the strong $l=2N$ peak leads us to consider odd frequency pairing below.

%Solving the gap equation for odd $l$ yields the standard BCS superconducting $T_c$
%and solve the gap equation. Therefore, in the regime of weak pairing interaction $g^{(l)}= \nu_0\Gamma^{(l)} \ll1 $ ($l \in {\rm odd}$), the superconducting $T_c$ is given by 

%where $g^{(l)} = \nu_0 \Gamma^{(l)}(0)  $.
%Here, the factor of $2$ in front of $g^{(l)}$ is due to the antisymmetrization of pairing interaction, which doubles the effective pairing interaction strength in odd-$l$ channel. 
 %as shown in in the quadratic Dirac model, and at $l=7,9$ in the quartic one 

%\addZ{In our analysis above, we find the even frequency pairing channels can only utilize the subleading partial wave component of pairing interaction as they feature odd angular momenta. This motivates us to consider odd frequency pairing, which features an even angular momentum, and thus can utilize the leading partial wave component of pairing interaction $l=2N$. However, it comes with a price that the odd-frequency channels do not have a BCS logarithm in its susceptibility \cite{RevModPhys.91.045005,berezinskii1974new}, and instability is expected to occur above a finite threshold of coupling strength. }

We conclude that our prediction of even-frequency pairing in even-$N$ Dirac model is solid, as it is safely in the controlled expansion regime. Nevertheless, we should note that this conclusion depends on the assumption of a bare interaction $V_q$ of the form given in Eq. \ref{eq:V(q)}. If there is substantial $q$ dependence on the scale of $k_F$, the first order process does not cancel and can be expected to be repulsive, which will tend to suppress pairing. A $q^2$ dependence will only affect the $l=1$ channel, so this effect will diminish as $l$ becomes larger.

%\addZ{\section{DMRG result on 1D two-band model}}

%\noindent\rule{\linewidth}{2pt}\par

%\par\addZ{\noindent\hrulefill}\par

%\par\noindent
%\addZ{
%\rule{8cm}{1pt} \\[-2ex]   
%\rule{8cm}{1pt}              
%}
%\par

\begin{figure}
    \centering
    \includegraphics[width=0.99\linewidth]{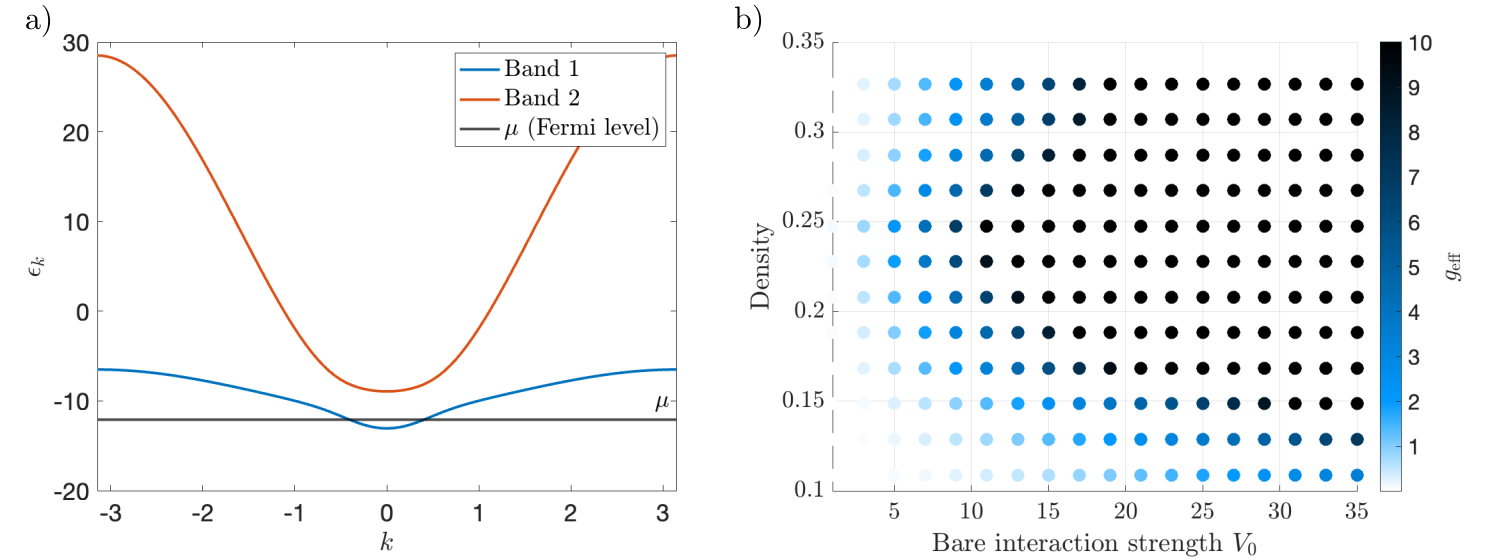}
    \caption{a) The band dispersion in 1D tight-binding toy model constructed for DMRG anlaysis. b) Effective coupling strength in p-wave Cooper channel in this model calculated from Feynman diagrams in Fig.\ref{fig:diagrams}.}
    \label{fig:band dispersion}
\end{figure}

\textit{ Two orbital tight binding model and DMRG evidence for p-wave pairing interaction:}
As a second example we consider a square lattice with two orbitals (A and B) which are located on the lattice and the center of the unit cell respectively. The orbitals can be both s or d. This simple model satisfies the requirement that the Hamiltonian $H(k)$ for the Bloch function $k$ is even in k. Hence the requirement $|u_{\vec k}\rangle = |u_{-\vec k}\rangle $ is satisfied and the first order term exactly cancels. Our strategy is to find a parameter range where $\alpha$ is less than unity but not too small, so that if we start with a strong but finite repulsion, the effective attraction is of order unity, giving high $T_c$. For simplicity we consider a 1D model, which has the additional advantage that our prediction can be accurately tested by DMRG. We consider the  following tight-binding Hamiltonian in momentum space:
\be
H(k) = \lp
\begin{matrix}
    \frac{u}{2} -2t_{AA}\cos k   & 2t_{AB} \cos \frac{k}{2}\\
    2t_{AB} \cos \frac{k}{2} & -\frac{u}{2} -2t_{BB}\cos k
\end{matrix}
\rp
\ee
To be concrete, we choose the following set of parameters: $t_{AA} = 10$, $t_{BB} = 1$, $t_{AB} = 1$, $u=17$, resulting in  the band dispersion shown in Fig.\ref{fig:band dispersion} a). The two bands hybridize near a small region at the center of Brillouin zone $|k|<k_*$. The interaction is given by the Theta function form as in Eq.\eqref{eq:V(q)}, with $k_0$ chosen to be 1.5. We focus on the regime of dilute carrier density, so that the Fermi surface lies in the hybridized region where $|u_k\rangle$ has a $k$ dependence. This model satisfies the requirement that $k_F$  is less than $k_0$ and the Bloch wavefunction has some variation near $k_F$ which is small but not too small. 
%This model is shown to have pairing by solving the BdG equation. 

As a preparation, before diving into DMRG we first map out the phase diagram of this 1D model using our analytic theory. 
In Fig.\ref{fig:band dispersion}b) we show the effective interaction $g_{\rm eff}$ as a function of interaction strength and carrier density. The $g_{\rm eff}$ is calculated as follows: First, diagonalize the kernal in BdG equation Eq.\eqref{eq:gap eqn 1} at a given temperature $T$. Here we limit ourselves to frequency-independent and spatially-odd channels. 
The eigenvectors are pairing channels, whereas we know from the standard solution of $T_c$ that the eigenvalues correspond to $g_{\rm eff} \log \frac{T}{W}$ for each channels. Therefore, we focus on the leading eigenvalue, do it for two slightly different values of $T$ and take numerical derivative over $\log T$ to extract the leading channel's effective coupling $g_{\rm eff}$. %computed from summing up first and second order diagrams in Fig.\ref{fig:diagrams}.
\footnote{In this calculation, for simplicity, we used the non-interacting band dispersion and Bloch wavefunctions, ignoring the Hartree-Fock renormalization of the band.} 

Our controlled expansion theory is applicable in the lower-left part of Fig.\ref{fig:band dispersion}b) where the density is sufficiently low so that $2k_F<k_0$ and the bare interaction strength can exceed order-1 but not too large $V_0 \lesssim O(10)$. In this regime, we indeed find an attractive effective interaction ($g_{\rm eff} >0$) as predicted by our   theory. This effective interaction increases with density and bare interaction and quickly reaches order-1. Further increasing the bare interaction, the system enters a strong-effective-coupling regime (black dots, $g_{\rm eff}\gg 1$). In that regime, our theory is no longer controlled.

When density exceeds $n=0.25$, the system enters a regime where our theory is not applicable. In this high-density regime, $2k_F$ exceeds $k_0$. This leads to the absence of backscattering, which invalidates the cancellation of first-order low-energy process in our theory. We expect no pairing in this regime because on Fermi surface, up to first order, there is only the forward scattering ($+k_F\rightarrow +k_F$, $-k_F\rightarrow -k_F$) which is pair-breaking. However, the result in Fig.\ref{fig:band dispersion}b) suggests that the leading $g_{\rm eff}$ is attractive in this regime (see blue dots in the upper left corner). This is confusing at first sight. To understand it, we need to look into the gap function in this leading attractive channel.
In 1D, $\Delta$ dependents on the distance away from the Fermi points $\pm k_F$.
We find the gap function is large away from the Fermi points, vanishes and changes sign precisely at the Fermi points, unlike the usual pairing model in which gap is finite on the Fermi level. Forming such a sign-changing gap across Fermi level is reasonable as this is the most natural way to avoid the repulsive forward scattering. The attraction in such channel can arise through the scattering between the gap near $k_F$ and the gap around momentum $\pm k_F\pm k_0$. As the latter momentum is away from Fermi level, the effect of such process is punished by a large denominator in the kernal (see Eq.\eqref{eq:gap eqn 1}). However, this punishment is not strong enough to suppress their effects as the dispersion is merely parabolic, unlike in the Dirac model where it diverges quickly as $k^N$ at large momentum. 
This is a result of the interaction far from Fermi surface, where the eigenvalue should not have a $\log T$ dependence. Our procedure of extracting $g_{\rm eff}$ assume a $\log T$ dependence in the susceptibility, which breaks down for these channels. Therefore, the $g_{\rm eff}$ we extracted here is no longer reliable and should not be taken seriously.
%This explains why we see such channels with a significant effective attraction in high-density regime. As these attractive channels do not gap out the Fermi surface, their presence is not tied to pairing.

%Yet, in Fig.\ref{fig:band dispersion} b), we see some attraction (blue dots) above the threshold density. However, a closer look confirms that this is not a real pairing, as the channel on which interaction's projection is attractive is not a real BCS pairing channel. Namely, we find the gap in attractive channels are all vanishing and sign-changing sign right at the Fermi level. In comparison, in this regime those normal p-wave pairing channels that opens a gap at the Fermi level feature a repulsive projected interaction. Therefore, we conclude that, even if the effective interaction is attractive when projected onto some channels, it does not really support a pairing in this large-density regime.

The phase diagram Fig.\ref{fig:band dispersion}b) based on our theory encourages us to use this  as a guide to search the parameter regime where  pairing occurs in accurate numerical solutions of the two orbital model using DMRG. The results are discussed next.
%We confirm this by solving the model using DMRG. 
%To further substantiate the theory and show its generality, below we try to show the tendency toward SC in a lattice model by numerically solving the ground state through DMRG\cite{itensor}. 
%It is tempting to study the Dirac model on a quasi-2D geometry using DMRG. However, given that our theory is only applicable to a dilute carrier setting, the maximal system width that DMRG can handle (which is usually $\lesssim 10$)is not large enough as compared to the value of $1/k_F$. Therefore, below we turn to a 1D model instead, which is ideal setting for DMRG. However,
It is difficult to directly probe an SC order in 1D because electrons in 1D  forms a Tomonaga-Luttinguer liquid, where all the correlation functions are power-law. Therefore, even if there is indeed a pairing interaction, there will be no long-range SC order. However, the presence of a pairing interaction is still detectable by measuring the exponent of SC correlator. Namely, in the theory of Luttinger liquid, the sign of the exponents of the SC, CDW and single-particle correlation functions are given as follows\cite{giamarchi2004quantum}:

\begin{itemize}
    \item SC correlator $\langle \Delta^\dagger (x)\Delta(0) \rangle$ $\sim |x|^{-\eta_\Delta}$,  $\eta_\Delta = 2/K$ %($\Delta = cc$) 
    \item Density-density correlator $\langle \rho(x)\rho(0) \rangle$: 
    \\
    1) uniform part:$\sim K|x|^{-2}$, \\
    2) oscillatory part: $\sim \cos(2k_Fx) |x|^{-\eta_{\rho}}$, $\eta_\rho = 2K$ 
    \item Single-particle correlator $\langle c(x)c^\dagger(0)\rangle$ $\sim |x|^{-\eta_c }$, $\eta_c = \frac{K+1/K}{2}$
\end{itemize}
where $K = \lp \frac{1+g_4-g_2}{1+g_4+g_2} \rp^{1/2}$. The dimensionless coupling $g_4$ is the strength of forward scattering between two left-moving or two right-moving carriers, $g_4 = \Gamma_{LL \rightarrow LL}= \Gamma_{RR \rightarrow RR}$, whereas $g_2$ is the interaction between opposite movers (or equivalently, the antisymmetrized pairing interaction):  $ g_2 = \Gamma_{LR \rightarrow LR}-\Gamma_{LR \rightarrow RL}$, which takes negative value when $p$-wave pairing interaction occurs\footnote{The sign convention here is such that $\Gamma$ takes positive values for a repulsion. }.  As a result, the value of $K$ directly reflects the sign of pairing interaction. Therefore, to probe the pairing interaction, we simply need to measure the exponent of SC correlator $-\eta_{\Delta}$: $\eta_{\Delta}<2$ indicates a predominant SC order over CDW and the presence of $p$-wave pairing interaction. 
%\addZ{[Is this paragraph necessary or too slow?]}

%With these preparations, below we start by constructing a 1D two-band toy model, and then show through DMRG that a pairing interaction can be achieved with a pure repulsion. We consider a wire with staggered AB sites with the following tight-binding Hamiltonian:

%We focus on the regime of dilute carrier density, so that the Fermi surface lies in the hybridized region where $|u_p\rangle$ has a $p$ dependence. According to our theory above, we expect it to enables a nonzero effective pairing interaction. Upon increasing bare interaction, we expect the ground state to exhibit a predominant SC correlation with $\eta_{SC}$ decreasing. Such ground state will remain stable even when bare interaction well exceeds order-1, until the \emph{effective} interaction reach order-1 where the perturbation theory in series of the \emph{effective} interaction breaks down.

%\begin{figure}
%    \centering
%    \includegraphics[width=1.0\linewidth]{analyticphasediagram.png}
%    \caption{}
%    \label{fig:placeholder}
%\end{figure}

\begin{figure*}
    \centering
    \includegraphics[width=0.99\linewidth]{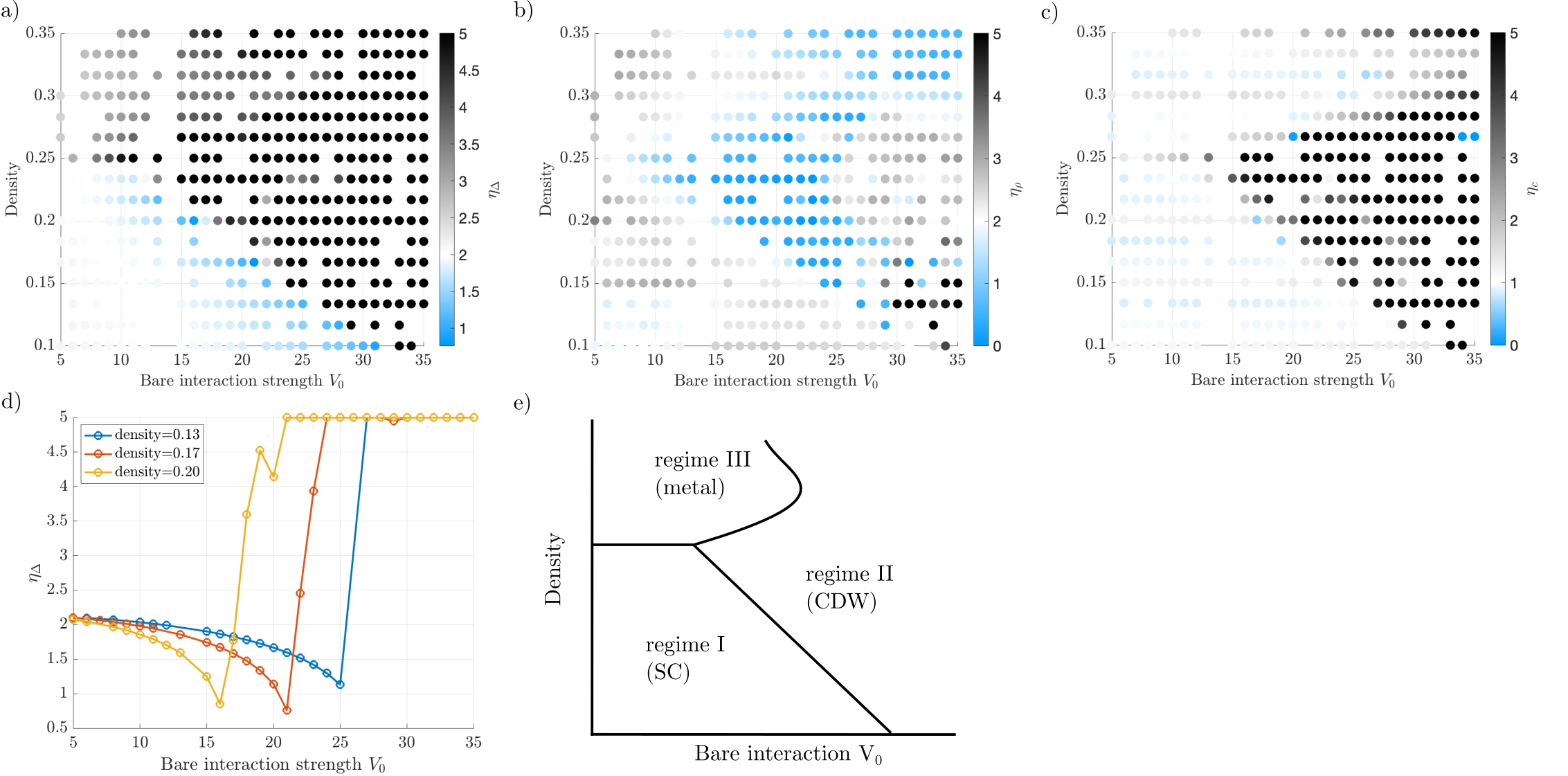}
    \caption{Exponents of three correlation functions a) $\eta_\Delta$, b)$\eta_\rho$, c) $\eta_c$, extracted from DMRG. Panel d) is the line cuts of panel a) along several values of densities, which shows the transition from SC regime to CDW regime is abrupt. Note that the pairing exponent $\eta_\Delta$ reaches the smallest value at the phase boundary with a value small enough to indicate strong pairing. (The values of the exponents in all four panels are cutoff between 0 and 5, which means all exponents exceeding 5 or below 0 have been set to the value of 5 or 0 respectively). Panel e): three regimes identified from these three exponents: I. low-density relatively-weak-interaction regime where SC correlation is dominant, II. strong-interaction regime where CDW order dominates, III. high-density regime (labeled as metal) where the system shows no tendency toward either SC or CDW order.
    %a) $\langle \Delta^\dagger(x)\Delta(0)\rangle$, b) $\langle \rho^\dagger(x)\rho(0)\rangle$, c) $\langle c(x)c^\dagger(0)\rangle$
    %. The blue dots represent the points where $\eta_{SC}<2$, which indicates a presence of pairing interaction. 
    }
    \label{fig:exponents}
\end{figure*}

Using DMRG, we obtain the three types of correlation function in ground state. The exponents are extracted in Fig.\ref{fig:exponents}. The $x$ axis is the interaction strength whereas the $y$ axis is the carrier density. These results show three regimes as described in caption. 

The formation of these regimes can be understood as follows:
Regime I (the lower-left corner) is the regime of dilute carrier density and a ``relatively weak interaction'' 
\footnote{We note that in regime I, the strength of interaction itself can be quite strong as the bare interaction can be pushed up to much greater than order-1. Here, we call it ``relatively weak'' merely because it is weaker compared to the regime II where, as we will discuss below, the interaction becomes even stronger strong so that the perturbation theory is no longer controllable. } 
which we are mainly interested in. In this regime,  the SC exponent $\eta_\Delta$ is below 2 whereas the density-density exponent $\eta_\rho$ is a little bit above 2. This matches the behavior of SC-dominated regime in Tomonaga-Luttinger theory. We note that the $\rho$ exponent here is measured by fitting the envelope of $\langle\rho(x)\rho(0)\rangle$ which contains both uniform component and $2k_F$ oscillatory component. Although the oscillatory part is expected to have an exponent $1/\eta_{\Delta}$ which is greater than $2$, the uniform part only has an exponent of 2. This explains why the measured density-density exponent only exceeds 2 by a little but never reaches $1/\eta_\Delta$. %Note that the value of $\eta_\Delta$ suggests the strength of pairing interaction ($g_2$ in TL theory) can be pushed up to order-1 at the transition from SC-dominated regime I to CDW-dominated regime II. 
%Meanwhile, These results match the predicted picture of our theory: at relatively weak interaction, the pairing interaction 

Further increasing interaction, our theory expects the controllability of perturbation theory to break down as the effective interaction  exceeds order-1. This predicted behavior is seen in the numerics: starting from SC-dominant regime and increasing interaction, we find the system abruptly enters regime II which shows a distinct behavior with $\eta_{\rho}<2$ and $\eta_\Delta>2$, implying a dominant CDW order. This transition is  discontinuous as the exponents changes abruptly at the boundary, as seen in Fig.\ref{fig:exponents} d). Interestingly the exponent decreases as the the phase boundary is approached from the SC-dominant side.  The value of $\eta_\Delta$ reaches the range 0.75 to 1.2,  suggesting that the strength of pairing interaction ($g_2$ in TL theory) can be pushed up to order-1 near the phase boundary. 

Upon further increasing interaction, more complicated behavior occurs where CDW order is suppressed again (see middle right part in Fig.\ref{fig:exponents} b)), but this behavior is beyond the scope of this paper as our perturbation theory already lost control under such strength of interaction. 

At a higher density, we expect a phase transition to occur when $2k_F$ exceeds $k_0$, which corresponds to a density of $\sim 0.25$ for the chosen value of $k_0$. At density above this threshold, the backscattering from $k_F$ to $-k_F$ becomes zero. As a result, the Tomonaga-Luttinger theory predicts $K=1$ and thus trivial values of exponents: $\eta_\Delta = \eta_\rho = 2$ and $\eta_c = 1$. Our numerics indeed match this expectation: Exactly at the expected threshold density, the system transition from SC-dominant regime (I) to a new regime (III) with the three exponents nearly taking trivial values.

%is abrupt, occurring at $n \approx 0.25$ almost independent of interaction strength. 
%This is reasonable because as analyzed above,  the interaction is so strong that the effective interaction exceed order 1 and perturbation theory breaks down.

%These results lead to the conclusion that the order-1 pairing interaction can indeed be achieved in a controllable manner like described in our theory. Moreover, we tested using other values of $k_0$ such as $k_0=2$ and $k_0=3$, the phase diagram remain similar except that the transition is pushed up to a higher density as expected. We also tested adding a power-law tail to $V(q)$ to mimic the momentum dependence of a realistic screened Coulomb interaction in gate-encapsulated geometry, which is constant at $k<k_0$ and $\sim k^{-1}$ if ignoring dielectric or $\sim k^{-2}$ if accounting for dielectric layer. We found the phase diagram does not change qualitatively. This shows the robustness of our theory.

A priori it is not at all obvious that a strongly repulsive two band model has a pairing regime  and it is not easy to find this without an exhaustive search.  These DMRG results demonstrate that our theory is indeed useful as a guide for us to reach a dominant $p$-wave pairing  with order-1 strength in the phase diagram.

We have performed further tests of the robustness of this conclusion. Namely, we tested other values of $k_0$, such as $k_0=2$ and $k_0=3$, where we find that the phase diagram remains qualitatively similar, except that the transition to regime III merely shifts to higher densities as expected. We also tried introduced a power-law tail to $V(q)$ to mimic the momentum dependence of a realistic screened Coulomb interaction in gate-encapsulated geometries—constant for $k<k_0$, scaling as $\sim k^{-1}$ without dielectric screening, or as $\sim k^{-2}$ when including dielectric effects\cite{Keldysh1979,Rytova2018,Cudazzo2011}. In both cases, the phase diagram showed no qualitative change. This robustness further supports the validity of our theory.

%\par\noindent
%\addZ{
%\rule{8cm}{1pt} \\[-2ex]   
%\rule{8cm}{1pt}              
%}
%\par

%To start, we construct a 1D two-band toy model that satisfy our recipe.

%re is still a crossover from "predominantly SC regime" to "predominantly CDW regime". The former occurs when the pairing interaction on Fermi surface is attractive whereas the latter occurs when it is repulsive. Usually, with an interaction $V(q)$ that monotonically decreases 
%If we can achieve a 
%in 1D wire, a fermionic system cannot 

%as the theory is only applicable to dilute carrier systems, applying DMRG to a 2D model is almost impossible.

To summarize, our new perturbative expansion relies on the following conditions: (1) a fully polarized band with low carrier density; (2) momentum dependence of Bloch wavefunction (3) a repulsion $V_q$ that is nearly $q$-independent at small q, which can be achieved through proximity to a screening metallic plane. %(3) inversion symmetry so that $|u_{\vec p}\rangle = |u_{-\vec p}\rangle $ , which ensures that the first order repulsive term is cancelled. Unlike the toy model, inversion symmetry is commonplace if the band is located at the zone center.
These requirements can be designed and  realized in various settings.   For example, in addition to gating, we can envision  layer by layer growth of ferromagnetic low carrier density  layers separated by conventional metals that act as screening planes, with a distance $d $ that can be nanometer or less, especially for van der Waals stacking. The ferromagnetism can be due to Stoner instability or exchange coupling to ferromagnetically aligned local moments. Low density carriers can be introduced by charge transfer from the metal. However, we caution that the two examples considered in this paper have the special feature that the Bloch function is even in momentum, i.e., $|u_{\vec p}\rangle = |u_{-\vec p}\rangle $. This leads to a complete cancellation of the first order term, leaving a second order term that is attractive. For general models, this condition is unlikely to be satisfied except for  special structures  and special assumptions about the orbitals so that the Hamiltonian for the Bloch function is  even in $\vec{k}$. In this paper we consider a Fermi surface near the zone center, but more generally the condition can apply to mmentum $p$ measured from a symmetry point in the Brilloiun zone, such as the zone corner. On the other hand, for a general Hamiltonian, we expect  the first order contribution to be significant. However, what is left after the approximate cancellation for the first and second order terms depends on the form factor $
\Lambda(\vec{k},\vec{k'})$ in very different ways and it is possible that the dominant repulsive channel in first order is different from the dominant attractive channel in second order. In that case pairing (or lack thereof) can be demonstrated in a controlled way. Such calculations will require detailed analyses of a given band structure and are beyond the scope of this paper.

The 1D example provides a stringent test of our model because there is a strong competing charge order instability in 1D which the superconductivity must overcome. This instability is in general not present in high dimensions, except for  Wigner crystal formation which requires very strong coupling, especially for a short range interaction. Therefore, the success of the 1D example gives us confidence for our formulation in higher dimensions. A second point is that since we start with a microscopic model, interaction with bands far away from the Fermi surface (Landau Fermi liquid effects) can renormalize the band parameters. This is why we consider a single Fermi pocket with low density, so that there are no occupied states to give a strong Hartree-Fock correction. Such effects can be further mitigated by considering models where the band disperses rapidly away from the Fermi surface. An example is the N-Dirac model where the band disperses as $k^N$ and interaction with large-momentum states have large energy denominators. Currently the models considered are developed mainly to illustrate the principle of achieving strong pairing in the presence of strong repulsion. We leave the possibility of application to real materials for the future.
%The leading interaction will likely have an attractive channel because second order perturbation theory is attractive. In this way one can search for topological superconductors with a relatively high $T_c.$

%We note at the end that a recent paper on pairing in a specific spin-polarized system \cite{May-Mann2025} appeared online just a few days before the first version of our work. That study also computed the pairing interaction up to second order, including all four diagrams, as in our approach. It is important, however, to distinguish the two works: The central contribution of our paper lies in pointing out the controlled regime of the usual perturbation theory can be greatly extended under certain generic conditions. 
%We identify a new small parameter that makes the perturbative expansion strictly controlled even in the regime of strong bare repulsion. 
%By contrast, Ref.\cite{May-Mann2025} only claim their calculation to be valid for weak-bare-interaction regime as usual. Addressing the controllability of perturbation theory in strong-coupling regime is not a goal of that study.}
Finally we note that upon the completion of this work, we became aware of a recent paper on pairing in spin-polarized system \cite{May-Mann2025} which also computed the pairing interaction in the $N$-Dirac model up to second order, including all four diagrams, as in our approach. However, the emphasis of that paper is on the effect of Berry curvature and the appearance of an additional small expansion parameter which is the main point of our paper was not addressed.

We thank Jason Alicea, Senthil Todadri, Andrey Chubukov, Ashvin Vishwanath, Leonid Levitov, Liang Fu, Zhihuan Dong, Tonghang Han and Jixiang Yang for insightful discussion. We thank Shengtao Jiang for guidance on using the ITensor package. Z. D. acknowledges support from the Gordon and Betty Moore Foundation’s EPiQS Initiative, Grant GBMF8682. P.A.L. acknowledges support from DOE (USA) office of Basic Sciences Grant No. DE-FG02-03ER46076.

\bibliography{ref}

\begin{thebibliography}{43}
\bibitem [{myf()}]{myfootnote1}%
  \BibitemOpen
  \bibinfo {note} {{In this figure, we used $\sqrt{g_2^{(l)}}$ as the variable
  in x-axis because this quantity scales linearly with $g_0$}}\BibitemShut
  {NoStop}%
\end{thebibliography}

\appendix

\section{Odd-frequency pairing}\label{app:odd_frequency}

\begin{figure}
    \centering
    \includegraphics[width=0.98\columnwidth]{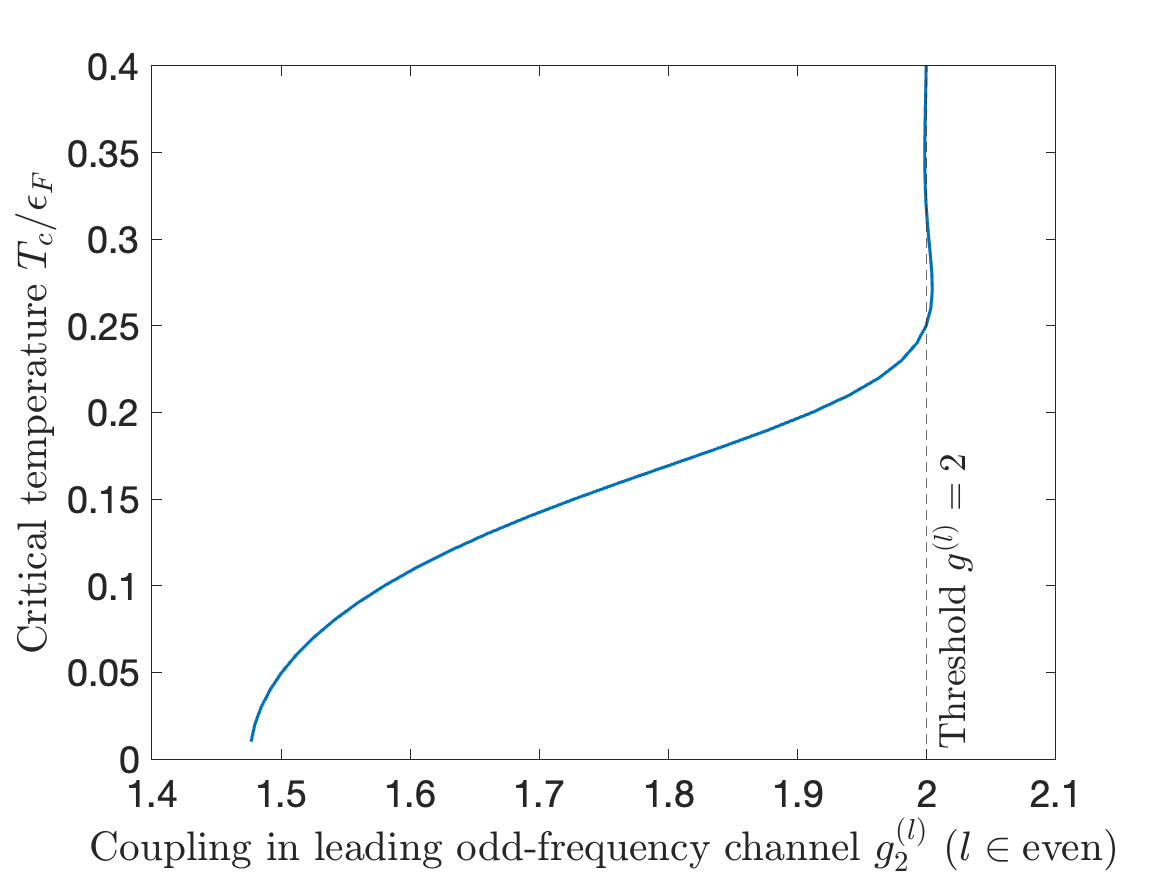}
    \caption{$T_c$ in odd-frequency channel as a function of coupling constant $g^{(l)}_{2}$. The behavior at $T_c\ll \epsilon_F$ agrees with analysis. The $T_c$ diverges at around $g^{(l)}_2=2$, but this is an artifact that arises due to neglecting $T$-dependence of pairing interaction in our analysis. We expect the $T_c$ to saturate somewhere $T\sim\epsilon_F $ when $T$-dependence of pairing interaction is accounted for(see text).}
    \label{fig:Tc odd frequency}
\end{figure}

%\addZ{In our analysis above, we find the even frequency pairing channels can only utilize the subleading partial wave component of pairing interaction as they feature odd angular momenta. This motivates us to consider odd frequency pairing, which features an even angular momentum, and thus can utilize the leading partial wave component of pairing interaction $l=2N$.}
In this appendix we analyze the odd-frequency pairing.
%Nevertheless, the strong $l=2N$ peak leads us to
The odd-frequency channels do not have a BCS logarithm in its susceptibility \cite{RevModPhys.91.045005,berezinskii1974new}, and instability is expected to occur above a finite threshold of coupling strength. To solve this threshold analytically, we model the frequency dependence of $\Gamma(\omega)$ using the following separable form:
%Therefore, we first study the onset condition of odd-frequency channels where Tc starts to becomes nonvanishing. This is essentially a zero-temperature problem, so we substitute Matsubara frequency summation $T\sum_{\omega'}$ with $\int \frac{d\omega'}{2\pi}$. To proceed, let us start by modeling (replacing) the $\Gamma_{\rm o}$ with an analytic function which a) mimics the realistic frequency dependence of $\Gamma_{\rm o}$, and at the same time b) keeps the equation analytically soluble. We note that the realistic frequency dependence of $\Gamma_{\rm o}$ should behave smoothly for $\omega\ll \epsilon_F$ and decrease to 0 for $\omega\gg \epsilon_F$, as it inherits the frequency dependence of particle-hole bubble.  We find the following form of $\Gamma$ can simultaneously satisfies these two requirements:
%To proceed analytically,  we model $\Gamma_{\rm o}(\omega-\omega';p,p')$ using a separable form:
\begin{align}\label{eq:Gamma separable form}
\Gamma^{(l)}(\omega-\omega') &= \Gamma_1^{(l)} + \Omega(\omega-\omega')\Gamma^{(l)}_2(0), \quad l\in {\rm even}\\ 
\Omega(\omega) 
%=\addZ{\frac1{\mathcal{N}}} \int d\tau \frac{\epsilon_F^2 \tau}{\rm{erfi}(\frac{\epsilon_F\tau}{\sqrt{2}}) } e^{i\omega\tau } 
&=\frac1{\mathcal{N}} \int \frac{\tau  e^{i\frac{\omega}{\epsilon_F}\tau }d\tau }{\rm{erfi}(\frac{\tau}{\sqrt{2}}) } , \nonumber
\end{align}
where ${\rm erfi}(x) = \frac{2}{\sqrt{\pi}} \int ^x_0 \exp(z^2) dz$, $\mathcal{N} = 4.6362$ is the normalization coefficient that makes $\Omega(0)=1$. Here, we isolated first-order contribution $\Gamma_1$ which is frequency-independent and therefore do not contribute to odd-frequency pairing. We chose this separable form %$\Omega(\omega)$ 
as a model for $\Gamma_2^{(l)}(\omega)$ because it mimics the realistic bandwidth of 2nd-order interaction ($\sim \epsilon_F$) and meanwhile keeps the gap equation analytically solvable.
Using Eq.\eqref{eq:Gamma separable form}, we find the following form of gap function is the odd-in-$\omega$ even-$l$ solution of $T=0$ linearized gap equation 
%[See Appendix \ref{app:check solution}] :
%separable-form solution: 
%\be\label{eq:Delta_o separable}
%\Delta_{\rm o} (\omega, p)= W(\omega) P(\theta), 
%\ee
%where the frequency dependent part takes the form of
\be\label{eq:T=0 solution}
%W(\omega) = 
\Delta^{(l)}(\omega) 
 = \omega \exp(-\omega^2/2\epsilon_F^2), \quad l\in {\rm even}
\ee
%Defining $g^{(l)}_{\rm o} = \nu_0 \Gamma^{(l)}_{\rm o}(0) $,
Plugging Eq.\eqref{eq:T=0 solution} into the $T=0$ gap equation yields: 
\be
1=\frac{\pi}{\mathcal{N}}g_2^{(l)}, \quad l\in {\rm even}
\ee
where $g_2^{(l)} = \nu_0 \Gamma_2^{(l)}(0) $.
This equation gives the onset condition for odd-frequency SC: $g_2^{(l)} = \mathcal{N}/\pi = 1.48 $. 
Once coupling $g_2^{(l)}$ exceeds this threshold, we expect that $T_c$ should quickly rise to $O(\epsilon_F)$ and saturates. 
%\addZ{We analyzed more carefully the behavior of $T_c$ within this model (Eq.\eqref{eq:Gamma separable form})
%, but we leave the details to
%Appendix~\ref{app:details of odd-frequency}}, as the results are approximate and depend on the specific form of $\Gamma(\omega)$ we chose. 
We note that odd frequency pairing is gapless because $\Delta(\omega=0) =0$, and is not expected to be topological, unlike the even frequency pairing.
So far, we have shown that the odd-frequency pairing sets in when the effective coupling $g_2^{(l)}$ exceeds an order-1 threshold value.
%, and expect that its $T_c$ can be comparable to $\epsilon_F$. 
%In this section,
Next, we calculate the dependence of $T_c$ in odd-frequency channels on $g^{(l)}_2$. For that, we restore the Matsubara summation in linearized gap equation and solve it numerically. When solving it, for simplicity, we first ignore the temperature dependence of pairing interaction $\Gamma$, which we will come back to comment on shortly. The numerical result is shown in Fig.\ref{fig:Tc odd frequency}. Here we see that SC indeed sets in when coupling strength $g^{(l)}_2$ exceeds a threshold of $1.48$
%0.738
and grows with $g^{(l)}_2$ as expected. 

However, $T_c$ behaves abnormally when it becomes $\sim \epsilon_F$: it abruptly diverges upon $g^{(l)}_2$ reaches a threshold of $2$. This behavior can be understood as follows: In the regime of $T_c\sim \epsilon_F$, the linearized gap equation becomes solvable again because $\Omega(\omega)$ takes nonzero values only at 
$\omega=0$, and therefore $\Delta(\omega)$ is nonzero only at the two smallest nonzero Matsubara frequencies $\omega = \pm 2\pi T_c$. In this case, the linearized gap equation becomes
\be
\Delta^{(l)}(\pm 2\pi T_c) = T_c\frac{\pi \Omega(0) g^{(l)}_2}{2\pi T_c} \Delta^{(l)}(\pm 2\pi T_c)
\ee
which yields a solution of $g^{(l)}_{2} = 2$, independent of $T_c$. It means when $g^{(l)}_{2}$ exceeds this threshold, SC can occur at any temperature, which is exactly the behavior seen in Fig.\ref{fig:Tc odd frequency}. 

However, we know that this behavior is nonphysical, as a temperature comparable to $\epsilon_F$ would suppress the susceptibility, thus suppress the pairing interaction. This T-dependence of pairing interaction is not accounted for in our calculation above. Therefore, we conclude that $T_c$ will saturate at $O(\epsilon_F)$.

%\section{Check the solution of linearized gap equation}\label{app:check solution}
In the end, we explicitly check that Eq.\eqref{eq:T=0 solution} is indeed the solution of Eq.\eqref{eq:gap equation angular momentum}.
Plugging Eq.\eqref{eq:T=0 solution} back in, focusing solely on the frequency dependent parts on both left and right hand side, and Fourier transforming from frequency domain to time domain, we find the Fourier transform of left-hand side in Eq.\eqref{eq:gap equation angular momentum} is
\begin{align}
\mathcal{F}\lb {\rm LHS} \rb &=
%=\mathcal{F}(W) = %T\sum_\omega 
\int \frac{d\omega}{2\pi} \omega \exp(-\omega^2/2\epsilon_F^2) e^{-i\omega\tau} 
\\
&= - i(2\pi)^{-1/2} \epsilon_F^3  \tau e^{-\tau^2 \epsilon_F^2/2},
\end{align}
whereas the Fourier transform of the right-hand side of Eq.\eqref{eq:gap equation angular momentum} is
\begin{widetext}
\begin{align}
    \mathcal{F}\lb \rm{RHS} \rb &= \pi g^{(l)}_2 \mathcal{F}\lb \Omega \rb \cdot\mathcal{F}\lb \frac{\Delta^{(l)}(\omega)}{|\omega|}\rb \nonumber\\
    &= \pi g^{(l)}_2 \lb \frac1{\mathcal{N}} \frac{\epsilon_F^2 \tau}{\rm{erfi}(\frac{\epsilon_F\tau}{\sqrt{2}}) } \rb \lb -i(2\pi)^{-1/2}  \epsilon_F \exp(- \epsilon_F^2 \tau^2/2) {\rm erfi}\lp \frac{\epsilon_F \tau}{\sqrt{2}}\rp \rb  \nonumber\\
    &=-i(2\pi)^{-1/2} \frac1{\mathcal{N}} \pi 
 g^{(l)}_2\epsilon_F^3\tau \exp(- \epsilon_F^2 \tau^2/2) 
    \end{align}
    \end{widetext}
Therefore, left-hand side and right -hand side indeed match when $
1=\frac{\pi}{\mathcal{N}}g^{(l)}_2$.

%However, the solution of $T_c$ in the regime of $T_c \gg \epsilon_F$ is not physical. It is an artifact due to ignoring the T-dependence of the pairing interaction $\Gamma(\omega)$ in our analysis. This approximation breaks down for $T_c\gtrsim \epsilon_F$ because the pairing interaction $\Gamma(\omega)$, which arises from second-order diagram and inherits the $T$-dependence of polarization function, should start to get suppressed when temperature becomes comparable to $\epsilon_F$, Therefore, even if our analytic solution gives an infinite $T_c$, we should still keep in mind that $T_c$ always saturates at $T_c\sim \epsilon_F$ in our model.

%\begin{widetext}

%\section{my convention}
%The interaction is given by
%\be
%H_{int} = \frac{1}{2}\sum_{\vec q} V(\vec q) :\rho_{-\vec q} \rho_{\vec q}:
%\ee
%The density operator is defined as $\rho_q = \int dr \psi^\dagger_r \psi_r e^{-iqr}$, where the field operator can be written in terms of Bloch wavefunction as
%$\psi_r = \sum_{k\in BZ} \Psi_k(r) \psi_k= \sum_{k\in BZ} u_k(r) e^{ikr} \psi_k $.  Therefore,
%\begin{align}
%\rho_q &= \sum_{kk'\in BZ} \langle u_{k'}| e^{i(k-k'-q)r}|u_k\rangle  \psi^\dagger_{k'} \psi_k \\
%&= \sum_{kk'\in BZ} \sum_G \delta_{G,k-k'-q}\langle u_{k'}| e^{iGr}|u_k\rangle  \psi^\dagger_{k'} \psi_k 
%\end{align}

%second order diagram
%\be
%\sum_{kk'}\langle p |  T_{kk'} | p' \rangle \langle -p |  T_{k'k} | -p' \rangle  \frac{f_k-f_{k'}}{\epsilon_k - \epsilon_{k'}}
%\ee
%\end{widetext}

\end{document}